\documentclass[pre,reprint]{revtex4-1}
\usepackage{subfigure}
\usepackage{graphicx}%
\usepackage{dcolumn}%
\usepackage{bm}%
\usepackage[english]{babel}
\usepackage{amsmath,amssymb}
\usepackage{upgreek}
\usepackage{esvect}
\usepackage{longtable}
\usepackage{epstopdf}
\usepackage{float}

\draft 

\begin{document}

\title{Influence of humidity on the tribo-electric charging and segregation in shaken granular media} 

\author{Andr\'e Schella}
\email[]{andre.schella@ds.mpg.de}
\author{Stephan Herminghaus}
\affiliation{MPI Dynamics and Self-Organization, Am Fassberg 17, 37077 G{\"o}ttingen, Germany}

\author{Matthias Schr\"oter}
\email[]{matthias.schroeter@fau.de}
\affiliation{Institute for Multiscale Simulation, N{\"a}gelsbachstrasse 49b, 91052 Erlangen, Germany}

\date{\today}
\begin{abstract}
We study the effect of humidity on the charge accumulation of polymer granulates 
shaken vertically  in a stainless steel container. The setup allows to control the humidity level from 5\,\% to 100\,\%RH while performing automated charge measurements in a Faraday cup directly connected to the shaking container. We find that samples of approximately 2000 polymer spheres become highly charged at low humidity levels ($<$ 30\%RH), but acquire almost no charge for humidity levels above 80 \%RH. The transition between these two regimes does depend on the material, as does the sign of the charge. For the latter we find a correlation with the contact angle of the polymer with only very hydrophilic particles attaining positive charges. We show that this humidity dependence of tribo-charging can be used to control segregation in shaken binary mixtures.
\end{abstract}


\maketitle 


\section{Introduction}\label{sec:introduction}

\begin{figure}[htbp]
\includegraphics[width=1\columnwidth]{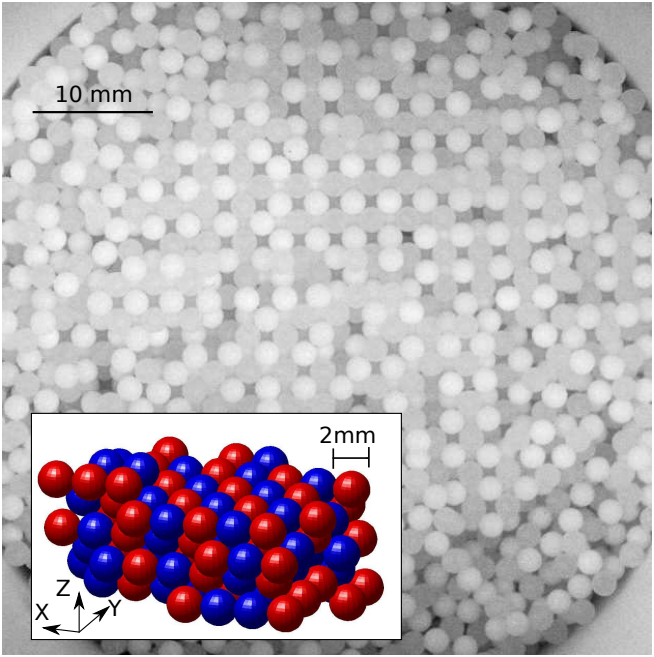}%
	 \caption{Crystalline structures formed by shaking PA and PTFE beads of 2\,mm size in a PTFE container at 50\,Hz and 2\,g for 1h.
 A body centered square lattice  (cesium chloride crystal structure) is visible in the upper part of the image. 
The inset shows a 3D reconstruction of the particle positions, obtained by X-ray tomography.
}
	 \label{fig:granular_crystal}%
\end{figure} 

Granular materials are among the most significant forms of matter in technology as well as in everyday life. Their production and handling, which accounts for about ten percent of the global energy consumption\cite{duran:00}, is hampered by a number of fundamental problems which are often not well understood. One of them is the effect of tribologic charging, which can occur whenever the grains are composed of a material different from the container, if two different grain materials are involved~\cite{Itakura1996,Nemeth2003,Nomura2003,Park2008,Liao2011,Matsusaka2010,Kumar2014} or even if grains differ in size~\cite{Kok2009,Xie2013,Waitukaitis2014}. Tribo-electric charges  can change the spatial composition of granular materials by inducing mixing~\cite{Lu2005,Cheng2014}, de-mixing~\cite{Mehrota2007,Forward2009}, clustering~\cite{Howell2001,Lee2015} or even the formation of granular structures~\cite{SaintJean2001,Grzybowski2003,McCarty2007,Kaufman2008,Chen2016} and crystals~\cite{Chademartiri2012,Lash2016,Qin2016}, cf. Fig~\ref{fig:granular_crystal}. Strong tribo-electric charging, which is capable of creating small sparks~\cite{Miura2007}, becomes even hazardous in the presence of flammable materials~\cite{Greason1992,Talawar2006,Eckhoff2009}. However, if charging is controlled it might provide a remedy for another well known granular problem: segregation. We show that indeed electrostatic attraction between particles of different size counteracts the mechanisms which normally make them separate when the sample is mechanically excited~\cite{ottino:00,Kudrolli2004,garzo:11}. Hence it is of great importance to understand, predict, and control tribo-electric charging in granular materials, in particular granular mixtures. 

Tribo-electric charging describes the ability of two bodies to exchange charges when getting in contact. The description of this phenomenon dates back to the ancient Greece~\cite{Ogrady2002}, but even nowadays it is an open debate whether electrons, ions or the exchange of surface material causes the net charge transfer from one contacting body to the other~\cite{Harper1957,Lowell1975,Duke1978,Lowell1980,Lowell1986b,Lee1994,Lacks2008,McCarty2008,Matsusaka2010,Lacks2011,Baytekin2011b,Galembeck2014,Xie2016}. This complicates the attempt to sort materials by their ability to accumulate charge when rubbed against each other~\cite{Shaw1917}. These so-called tribo-electric series are often mutually inconsistent and depend on hard to control details of the charge acquisition procedure~\cite{Diaz2004,Park2008,Burgo2016}. 

A number of experimental and theoretical studies suggest that water molecules adhered to the surface of the materials play an important role in the charge transfer~\cite{Harper1957,Pence1994,Greason2000,Howell2001,Nomura2003,Nemeth2003,Wiles2004,Diaz2004,Albrecht2006,Park2008,Kudin2008,McCarty2008,Albrecht2009,Gouveia2009,Ducati2010,Burgo2011,Galembeck2014,Shinbrot2014,Zhang2015,Xie2016}. Such water layers are ubiquitous, however their thickness is limited to the order of a few nm~\cite{Nemeth2003,Zhang2015} as long as the ambient humidity is below 100\,\%RH~\cite{Zitzler2002,herminghaus_wet_2013}. Already in 1902, Knoblauch hypothesized that the $H^{+}$ and $OH^{-}$ ions dissolved in the water adsorbed to the surface of polar solids would be reasonable charge carriers~\cite{Knoblauch1902}. Subsequent experimental observations confirmed the idea of the importance of surface water, however a consistent picture of its role still has to emerge. For instance, it was found that humidity strongly alters the charges that are generated via tribo-electricity~\cite{Nieh1988,Pence1994,Diaz2004,Wiles2004,Xie2016}. This is attributed to the effect that higher air humidities will increase the air conductivity~\cite{Zavattoni2013} and hence {\it increase the leakage} of charges from the surfaces to the ambient air~\cite{Harper1957,Nomura2003,Diaz2004,Burgo2011}. An {\it increased charging rate} for increasing humidity was found by Wiles \textit{et al.} who studied the charge transfer of a rolling metal sphere on a flat polystyrene surface~\cite{Wiles2004}. N{\'e}meth and coworkers concluded from their study of fluidized beds filled with polymer particles that the charge transfer mechanism is dominated by electrons at low humidity; at higher humidity adhered water and ions contribute to the electrostatic charging~\cite{Nemeth2003}. In other work, it was pointed out that water is not a necessity for the charge exchange~\cite{Baytekin2011a} 
or that even small patches of water together with large electric fields can lead to the charge transfer between solid, hydrophobic objects and walls\cite{Kok2008,Paehtz2010,Zhang2015}. 
Another mechanism of tribo-electric charging might be the exchange  of functional groups of the polymers, which are transfered to the surface of the beads when they are swelling at high humidities~\cite{Nemeth2003,Burgo2011}. 

To summarize, there is as yet no complete picture of the role of humidity on the tribo-electric charging. In the present study, we therefore concentrate on phenomenological parameters which can be determined experimentally, such as the wettability of the materials by water (i.e., the contact angle \cite{herminghaus_wet_2013}) and the relative humidity (RH). We control the humidity of the ambient air in a range from 5 to 100 \%RH to test the influence of this parameter on the charge accumulation of polymer spheres shaken vertically in a stainless steel container. 

\section{Experiment}\label{sec:experiment}
\begin{figure}[htbp]
\includegraphics[width=1\columnwidth]{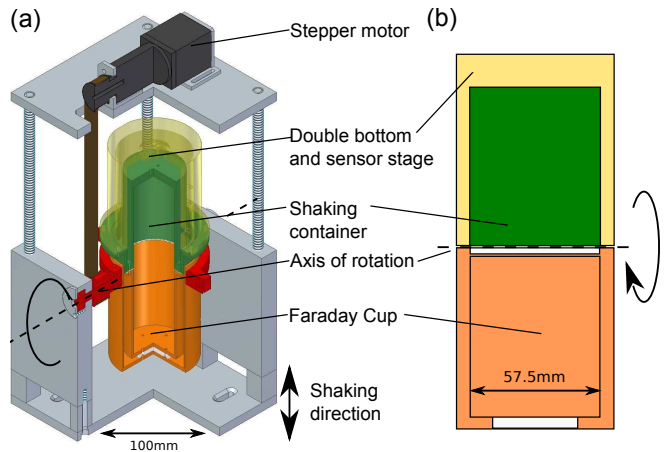}%
	 \caption{Rotating cup setup used for the automated charge measurements. (a) in an isometric - view and (b) in a simplified cross-section. Different structural components are color-coded in the same color in both views. The Faraday Cup (orange) is mechanically connected to the shaking container but electrically isolated by a 1\,mm PTFE ring (white). This assembly can be rotated around a common axis (red) using a toothed belt (brown) and a stepper motor (black). The shaking container (green) has 40 1\,mm pin holes which allow an exchange of humidity with the outer hull (yellow). The outer hull itself can be flushed with air of controlled humidity via external valves; it also contains a temperature and a humidity sensor. The whole setup is mounted on top  of an electromagnetic shaker operated in vertical direction. Shown is the charge measurement position. }%
	 \label{fig:rotating_cup}%
\end{figure}

Most charge measurements involving granular materials use a Faraday cup, which implies that the sample has to be transfered from the point where it acquires its charge to the point where this charge is being measured~\cite{Greason2000,Nemeth2003,Liao2011,Soh2014}. Because this transfer process can influence the charge in an uncontrolled way, we have designed a setup where the Faraday cup is directly connected to the shaking container; the transfer of the beads is then initiated by rotating this assembly with a stepper motor, cf. Fig. \ref{fig:rotating_cup}. Both Faraday cup and shaking container are made from stainless steel (alloy 4301) and have an inner diameter of 5.75 cm. The whole setup is mounted on an electromagnetic shaker (TIRAvib TV 5880/LS) and shaken sinusoidally in a  vertical direction. The granular samples used for charge measurements consist of different types of polymer beads with a diameter of 3 mm. Table~\ref{tbl:materials}  lists the materials studied, together with the polydispersity of the beads and the literature values of the equilibrium contact angle $\Theta_{\rm Y}$ for water.  Except for the PTFE beads, which were obtained from TIS W{\"a}lzk{\"o}rpertechnologie~\cite{TIS_web}, all beads were ordered from Spherotech~\cite{Spherotech_web}. 

\begin{table*}[htbp]
  \caption{\ List of all polymer materials of the sphere in our experiments. The polydispersity is given as quoted by the distributor. The contact angle $\Theta_{\rm Y}$ (at water-air interface) are obtained from literature.}

\small
  \label{tbl:materials}
  \begin{tabular*}{0.5\textwidth}{@{\extracolsep{\fill}}llcc}
    \hline
    material & diameter& poly- & contact \\
		 & [mm] & dispersity & angle $\Theta_{\rm Y}$ ($^{\circ}$) \\
    \hline
    HDPE (high density polyethylene) & 3.0  & 1.7\,\% & 97~\cite{Ammosova2015}\\
    PA  (polyamide, Nylon 6-6) & 3.0 & 1.7\,\% & 70~\cite{Kwok1999} \\
    PA  &  2.0                      & 2.5\,\% & \\
    PA  &   1.59                  & 3.2\,\% \\
    POM (polyoxymethylen) & 3.0 & 0.8\,\% & 62~\cite{Loh1987} \\
    PMMA (polymethylmetharcylate) & 3.0 & 4.2\,\% & 73~\cite{Winters1985}\\
    PP (polypropylene) & 3.0  & 1.7\,\% & 100~\cite{Winters1985}\\
    PS (polystyrene) & 3.0  & 1.7\,\% & 91~\cite{Kwok1999}\\
    PTFE (polytetrafluoroethylene) & 3.0 & 0.8\,\% & 108~\cite{Kwok1999}\\
    PTFE &              2.0        & 2.5\,\% & \\
    PTFE  & 1.59                      & 3.1\,\% & \\
    PVC (polyvinylcloride) & 3.0  & 1.7\,\% & 80~\cite{Kwok1999}\\
    \hline
  \end{tabular*}
\end{table*}

\subsection{Measurement protocol}\label{subsec:measurement_protocol}
Prior to every measurement series, the particles were cleaned with de-ionized water and ethanol, and the Faraday cup and shaking container were wiped with ethanol. Subsequently, the shaking container was filled with either a single bead or with $2000\pm17$ particles. The latter corresponds to a filling height of approximately 2 cm, which is one fifth of the containers' height. 

The beads were shaken for at least 8\,min at an amplitude of 1.4\,mm and a frequency of 30\,Hz, which corresponds to an acceleration of 2.5\,g (g = 9.81\,$\mbox{m}/\mbox{s}^2$). 
Then the Faraday cup and container are rotated by 180$^{\circ}$ around a horizontal axis. After the rotation, an additional short shaking pulse of  1\,s ensures that even beads which are still sticking to the walls of the container fall in the Faraday cup.

The charge of the granular sample is measured using a Keithley 6514 electrometer connected to the Faraday cup. The shaking container is grounded and the charge is measured as the average over 50\,s. An additional drift correction of the electrometer was applied to all measurements. Uncertainty for the single bead measurements was $6.7\textsf{x}10^{-11}\,\mbox{C}$. 
After finishing the charge measurement, both Faraday Cup and shaking container are grounded. Then, the system is rotated back to its original position and a new shaking and measurement cycle begins. 

\subsection{Humidity control}\label{subsec:humid_control}
In order to tune the humidity within the setup, the shaking container is embedded in an outer hull. Forty pin holes in the container (diameter: 1\,mm) enable the exchange of air between the inside of the rotating cup setup and the space in the outer hull. To adjust the humidity in the range between 20 and 100\,\% RH, the outer hull was flushed with preconditioned air from a  self-built climate chamber which uses an ultrasonic transducer and a cold trap to control the humidity level. Measurements at a humidity level of 3-5\,\%RH were performed by fixing a bag filled with 12\,g dry Silica gel inside the outer hull. 

For monitoring purposes the outer hull is equipped with a humidity sensor (HIH-5030/5031 with an accuracy of $\pm3$\,\% RH in the range of 11-89\,\% RH and $\pm7$\,\% RH otherwise) 
and temperature sensors (TMP102, with an accuracy of  $\pm0.5^{\circ}$C); both sensor were purchased from Tinkerforge~\cite{Tinkerforge_web}.

Before starting the experiments, all air connections between the setup and the climate chamber were closed in order to maintain a closed atmosphere. The time scale for equilibration 
between the inner part of the sample chamber and the volume in the outer hull was found to be on the order of 80\,min. Additionally, the shaker produces heat during operation, which increases the temperature in the sample chamber till it reaches an equilibrium at $T\,=\,28\pm1\,^{\circ}\mbox{C}$ after approximately an hour. To account for both of this effects, we have only considered charge measurements where the  inner and outer part of the setup can be considered equilibrated. Finally, a drift of $\approx$ 0.3\,\% RH per hour was found for a humidity difference of 55 \% RH between the lab and the interior of the container. However,  all measurements are attributed to the present humidity level determined with the humidity sensor.

\subsection{X-ray micro-tomography}\label{subsec:tomography}
We studied segregation in binary mixtures using X-ray computed tomography. The setup (Nanotom, General Electrics) uses a tungsten target and was operated at 130\,kV acceleration voltage and 90\,$\mu$A current to produce X-rays. The data sets consist of $900\times900\times800$ voxels where each voxel has a size of 60$\mu$m$^3$. Air was removed prior to bead detection using a threshold determined via the Ostu method~\cite{Otsu1975}. The same approach was used to discriminate beads of different materials. After an image erosion step~\cite{Haralick1987,BurgerBurge_imageproc_2008} and binarization, the center of the beads was detected with an accuracy of 1\,\% of the small bead diameters, which is smaller than the polydispersity of the grains. 

\section{Results}\label{sec:results}
The aim of our work is to quantify the influence of humidity on the generation of charges in granular media. Section \ref{subsec:impact_humidity} describes the influence of humidity on shaken samples of four different granular materials. A possible relationship between tribo-charging and the wetting behavior of the polymer materials is discussed in Sec.~\ref{subsec:compare_wetting}. Section \ref{subsec:comp_single_multi} compares the different impact of humidity on bead-wall and bead-bead contacts. Finally, the role of humidity in the segregation of binary mixtures is highlighted in Sec.~\ref{subsec:segregation}. 

\subsection{Impact of the humidity level on the charge}\label{subsec:impact_humidity}
 Figure \ref{fig:impact_humid} depicts the charge of shaken samples of $N$ = 2000 beads made from PA, POM, PS or PTFE over the entire range of humidity. Three observations can be made from Fig.~\ref{fig:impact_humid}: 

\begin{figure}[htbp]
\includegraphics[width=1\columnwidth]{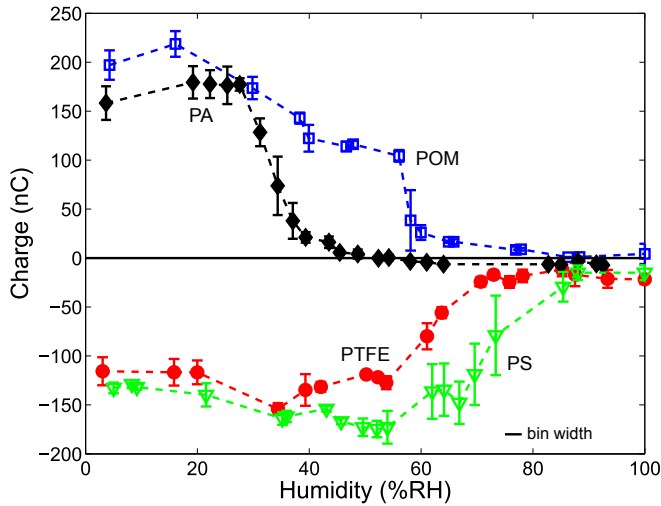}%
	 \caption{Magnitude of charge as a function of humidity for each 2000 PA, POM, PS, or PTFE spheres shaken in a steel container. Up to moderate humidities, all fillings are highly charged. Then, at a material specific humidity, a crossover to a low-charged regime is found. At very high humidities, only the hydrophobic materials PS and PTFE are able to accumulate charges. The data points correspond to bins of 3\% RH width, each containing at least three individual measurements.}
	 \label{fig:impact_humid}
\end{figure}

I) All materials are strongly charged at low humidities. At a material-specific  critical humidity  RH$_{\rm crit}$ there is a crossover into a low charge regime. We determine RH$_{\rm crit}$ by modeling the data in figure \ref{fig:impact_humid} with a 5th order polynomial and then computing the point where the negative slope has a maximum. RH$_{\rm crit}$ of all four materials is listed in table~\ref{tbl:charges}. 

II) The hydrophilic PA and POM beads are positively charged and the hydrophobic PS and PTFE beads are negatively charged. We will return to the sign of the charge in Sec.~\ref{subsec:compare_wetting}. 

III) The hydrophobic materials are still charged up to a few tens of nC even at very high humidities ($>80\,$\%RH). 

\begin{table}[htbp]
\small
  \caption{Electrostatic charge acquired by 2000 spheres after shaking in a steel container at low humidity ($\mbox{RH}<\,5\%$) together with literature value for the humidity at which a monolayer (ML) of water forms on the polymer surface, and the value RH$_{\rm crit}$ where the charge values presented in figure \ref{fig:impact_humid} cross over from large to small.}
  \label{tbl:charges}
  \begin{tabular*}{0.5\textwidth}{@{\extracolsep{\fill}}lccc}
    \hline
    material & charge  [nC] &  RH & RH$_{\rm crit}$\\
                 &               at $\mbox{RH}<\,5\%$                              &  at one ML of water     &                   \\
    \hline
    PA & $158.3\,\pm17.1$ & 10 \% \cite{Nemeth2003} & 28 \%  \\
    POM & $197.1\,\pm15.0$ &  n.n. & 46 \% \\
    PS & $-136.2\,\pm5.1$ & $>70$ \% \cite{Nemeth2003} & 78 \% \\
    PTFE & $-115.7\,\pm14.3$ & $\approx80$ \%  \cite{Zhang2015} & 61 \% \\
    \hline
  \end{tabular*}
\end{table}

Several authors observed changes in the magnitude of charge upon altering the ambient humidity with the general trend being that charging is suppressed at high humidity~\cite{Nieh1988,Pence1994,Greason2000,Nemeth2003,Xie2016}. However, there exists no unified \textit{ab-initio} model that directly relates tribo-charging to the thickness $h$ of the adsorbed water film. The exact value of $h$ depends on the effective interface potential of the surface and it increases in a nonlinear way with the relative humidity. For the case of incomplete wetted materials such as polymers the range of $h$ is between fractions to several tens of monolayers of water~\cite{Nemeth2003,Zitzler2002,herminghaus_wet_2013,Zhang2015}. An interpretation of our data has therefore to rely on a combination of the following mechanisms: 

First, if charging is due to the presence of water on the surface of the particles, free charge carriers are either supplied due to dissociation of water molecules present in the ambient air~\cite{Gouveia2009,Burgo2011,Ducati2010,Galembeck2014} or it could be ions transferred via "water bridges" from one body to the other upon contact~\cite{Pence1994,Kudin2008,McCarty2008,Zhang2015,Burgo2016,Xie2016}. Second, charging could be completely independent of water~\cite{Duke1978,Lee1994,Lacks2008,Baytekin2011a} and third, discharging can occur whenever the presence of surface water decreases the surface resistivity \cite{Nieh1988,Pence1994,Nemeth2003}. A possible fourth mechanism is the charge loss due to the increased conductivity of humid air~\cite{Carlon1981,Zavattoni2013}. However, this mechanism can at best explain differences between positively and negatively charged surfaces \cite{Burgo2011}, but not the observed material-dependent variations. We therefore exclude it from our further considerations.

The fact that we observe the strongest charge accumulation at the lowest humidity levels does not contradict the relevance of water for charging as suggested by the first mechanism, since charging will require only a small amount of water molecules to be present at the surfaces. However, it does also not exclude the possibility that other charge transfer mechanisms related to the second argument are responsible for the observed charging. 

The crossover to smaller or even zero charge at RH$_{\rm crit}$ can be explained by the third mechanism, an increased charge loss due to surface water~\cite{Nieh1988,Nomura2003}. Table~\ref{tbl:charges} lists some literature values for the formation of the first monolayer on polymer surfaces, which corresponds to a conducting hull around each particle; these values are in reasonable agreement with our measured values of RH$_{\rm crit}$. Moreover, Nemeth \textit{et al.}~\cite{Nemeth2003} measured the surface resistivity $\rho$ of PA (PA-12), POM and PS particles. For PA particles they found an approximately sevenfold decrease of $\rho$ between 30 and 50 \%RH. For PS particles there is a twentyfold decrease between 50 and 70 \%RH whereas the surface resistivity values of PS do not decrease up to a humidity level of 70 \%RH. All three of these results are in good agreement with our measured RH$_{\rm crit}$.

Finally, observation III, the finite charge on PS and PTFE beads at high humidity levels, supports the idea that this charge is related to water adhered to the surface. 

\subsection{Contact angle and wetting behavior}\label{subsec:compare_wetting}
\begin{figure}[htbp]
	\centering
\includegraphics[width=1\columnwidth]{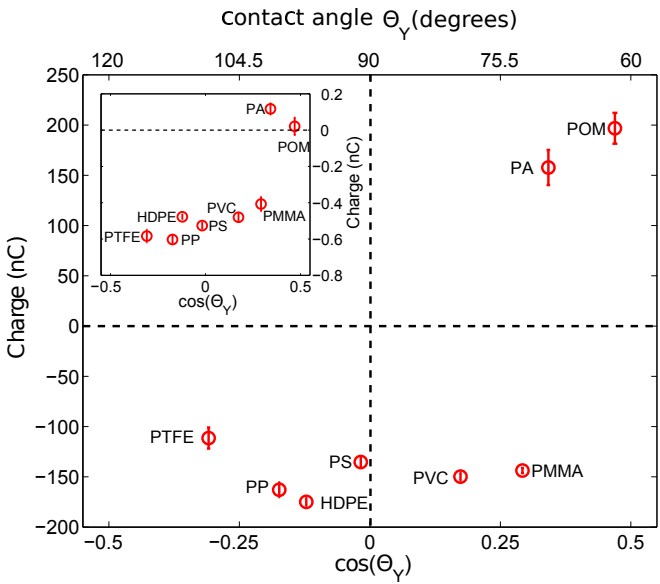}%
	 \caption{Charge of 2000 polymer beads shaken at $<$5\%RH versus the cosine of the contact angle  $\Theta_{\rm Y}$. $\Theta_{\rm Y}$ values are taken from literature according to table \ref{tbl:materials}. The inset depicts the charge of a single bead shaken in the same steel container at otherwise identical conditions. At least 9 (25) measurements were averaged for 2000 (a single) particle measurements. } %
	 \label{fig:cont_vs_char}%
\end{figure}

The data in figure \ref{fig:impact_humid} seem to imply that hydrophobic surfaces, i.e.~with a contact angle $\Theta_{\rm Y} > 90^{\circ}$ charge negatively while hydrophilic materials charge positively. However, this picture is oversimplified as the inclusion of a larger variety of polymer materials in figure \ref{fig:cont_vs_char} demonstrates: already for $\Theta_{\rm Y} > 70^{\circ}$, all samples are negatively charged. Only the the two most hydrophilic polymers, PA and POM, charge positively.

This more complicated dependence can be expected from the fact that there is no simple relation between the thickness $h$ of the adsorbed water film and $\Theta_{\rm Y}$ \cite{herminghaus_wet_2013}. Still there seems to be a correlation between $\cos(\Theta_{\rm Y})$ and the sign of the accumulated charge. This raises the question of the nature of the charge carriers that cause tribo-charging. Numerical results imply that hydrophobic graphene surfaces favor an adsorption of negatively charged hydroxide ions $OH^{-}$ from ambient air~\cite{Kudin2008}. Experimental support for this interpretation comes from experiments where water is flowing through PS and PTFE tubes and charges them negatively \cite{Burgo2016}. Clint and Dunstan~\cite{Clint2001} directly relate the wetting properties of materials with the capability of a surface to donate electrons and their position in the tribo-electric series. However, this does not exclude ions as charge carriers due to adhered surface water: It was observed that for instance metals tend to provide free charge carriers via ion partitioning at the solid-gas interface~\cite{Ducati2010} and that the tribo-charge does depend on details of the surface chemistry~\cite{Nemeth2003,McCarty2008} with an $H^+$-adsorption on basic and an $OH^{-}$-adsorption on acidic surfaces~\cite{Gouveia2009,Gouveia2012,Galembeck2014}. Xie \textit{et al.} proposed a surface state model with $H^+$-ions as good candidates for the donated charge transferred during a collision event between unequal sized glass spheres~\cite{Xie2016}. To summarize, there is evidence that the surface chemistry links both the wetting properties and the charging behaviour of granular materials. 

\subsection{Comparing particle-wall and particle-particle contacts}\label{subsec:comp_single_multi}
When shaking 2000 polymer beads in our steel container, a bead-bead collision is approximately nine times more likely than a bead container collision (based on the ratio of the total bead surface area of 565.5 cm$^2$ to the inner container area of 62.1 cm$^2$). Hence the charges presented in the last two subsections will be strongly affected by polymer-polymer collisions. However, due to charge conservation within the bed, the main source of charge are due to collisions with the container wall. 
Now to probe exclusively these polymer-steel collisions we have repeated the experiments with a single polymer sphere shaken in the steel container.

\begin{figure}[tbp]
	\centering
\includegraphics[width=1\columnwidth]{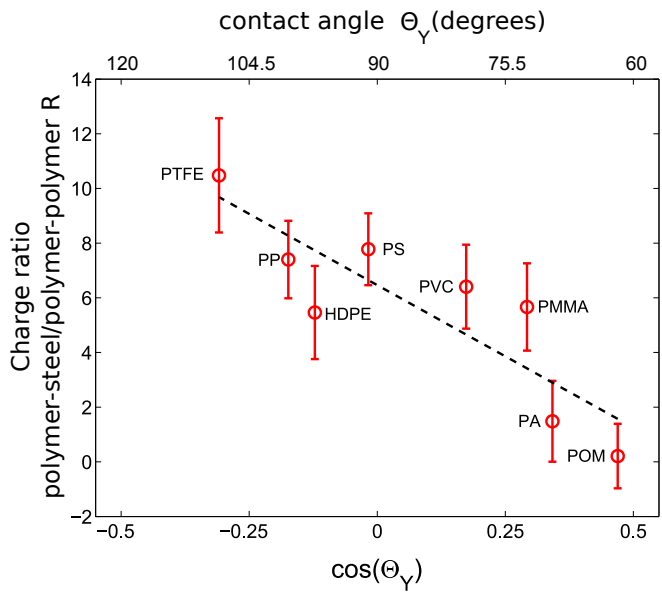}%
	 \caption{The ratio $R$ of the charges accumulated by a single sphere shaken in a steel container and the average charge of a bead shaken within a sample of 2000 particles; both shaken at dry conditions ($<5$\%RH). While hydrophobic particles charge stronger in polymer-steel collisions than in polymer-polymer collisions, this ratio decreases with with decreasing $\Theta_{\rm Y}$. Error bars were computed using error propagation and the data in Fig.~\ref{fig:cont_vs_char}. The dashed line is a guide to the eye. } %
	 \label{fig:comp_single_multi}%
\end{figure}

The inset of figure \ref{fig:cont_vs_char} shows that the correlation between the sign of the charge and $\Theta_{\rm Y}$ does not change for the polymer-steel collisions. However, the
amount of accumulated charge does. Figure \ref{fig:comp_single_multi} displays the ratio $R$ between the charge acquired by a single shaken bead and the charge acquired by a bead shaken within a sample of 2000 beads. Or, in other words, the charge ratio of polymer-steel to polymer-polymer collisions.

Figure \ref{fig:comp_single_multi} shows that $R$ decreases with decreasing contact angle. 

In general, metal-insulator contacts can result either in electron~\cite{Nemeth2003,Albrecht2009} or ion transfer processes~\cite{McCarty2008,Wiles2004,Diaz1993,Diaz2004}. Without a microscopic understanding of our charging mechanism it is difficult to explain the trend in figure \ref{fig:comp_single_multi}. However, the presence of water seems to be crucial for our observation as the observed hierarchy of charging is not compatible with measurements performed at polymer-metal contacts under vacuum conditions i.e. in the absence of surface water \cite{Diaz2004}. Moreover, the contact angle of stainless steel with water is roughly 66$^{\circ}$~\cite{Kalin2014}. Therefore the more hydrophobic a polymer is, the stronger will be the contrast in $\Theta_{\rm Y}$ and consequentially the exchange of water when the polymer gets into contact with steel. 

Collisions between the charged polymer particles can cause nonlinear effects~\cite{Siu2015,Qin2016} , diffusion of charges along the grain surface and contact de-electrification~\cite{Soh2014}. The latter mechanism according to Soh \textit{et al.}~\cite{Soh2014} implies that charges are transferred to the atmosphere upon contact. We can only say that contact de-electrification seems to be a stronger effect for smaller water surface layer thickness \textit{resp.} higher contact angles. This could be related to the Paschen-breakdown mechanism in the vicinity of the contact area between beads~\cite{Laurentie2013,Soh2014}.
\subsection{Using humidity to control segregation in binary mixtures}\label{subsec:segregation} 

\begin{figure}[htbp]
\includegraphics[width=1\columnwidth]{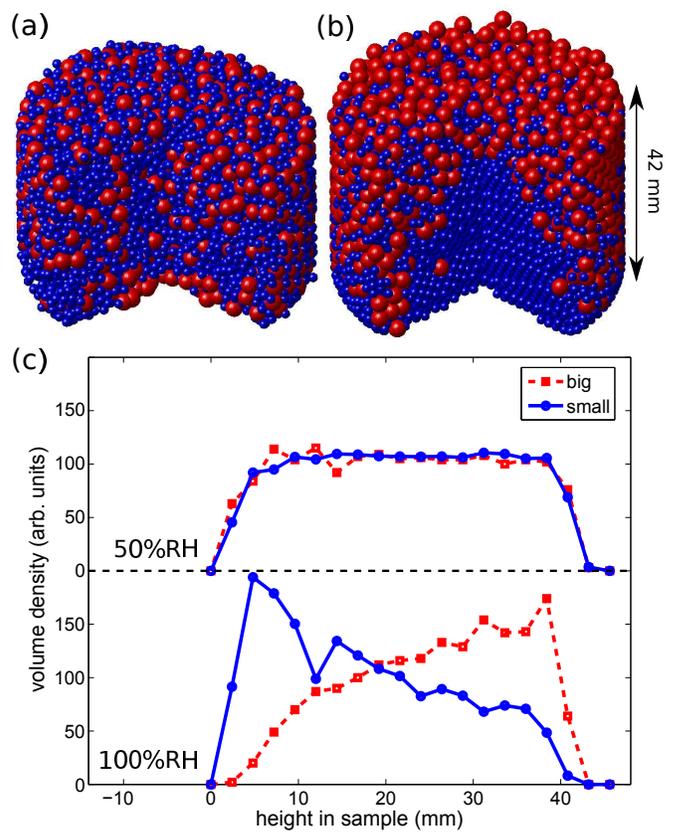}%
	 \caption{X-ray tomography demonstrates that segregation can be suppressed by tribo-electric charging and therefore controlled by the ambient humidity. {\bf a)} Rendering of an equal volume mixture of 1.59 mm and 3 mm diameter PTFE spheres shaken for one hour in a PA container at an ambient humidity of 50 \%. A wedge of 90$^{\circ}$ has been cut out for better visibility. The sample is well mixed. {\bf b)} The same sample as in a), but shaken at 100\%RH, is well segregated. {\bf c)} The height resolved volume density indicates the amount of segregation.}
	 \label{fig:density_tomos}%
\end{figure}

The fact that the ambient humidity determines the charge on the particles allows for some control of segregation in vertically shaken samples. To demonstrate this point we have mixed equal volumes of large (3\,mm) and small (1.59\,mm) PTFE beads and shaken them in a PA container (diameter 50\,mm) for one hour and at 100\,Hz and 2\,g. We then obtained X-ray tomographies of the samples and detected the particle positions. Figure \ref{fig:density_tomos} a) and b) depict the difference between samples shaken at ambient humidities of 50\%RH and 100\%RH. For 50\,\%RH, small and large particles are well-mixed within the sample. Fig.~\ref{fig:granular_crystal} shows that even crystalline mixtures may result when same-sized grains are shaken. This is most readily interpreted as being due to attractive electrostatic interactions between the beads. In contrast, at 100\,\%RH the sample displays horizontal and radial segregation, cf. figure~\ref{fig:density_tomos} c). This so called Brazil Nut segregation pattern can be explained by the two segregation mechanism of void filling and convection rolls \cite{Rosato1987,knight:93,Schroeter2006}. 

To study the degree of segregation in our binary mixtures, we define the dimensionless segregation parameter $P_{\rm segr}$ as 
\begin{equation}\label{eq:segregation_parameter}
	P_{\rm segr}= \frac{1}{2} \sum_{i = 1} ^{h_{\rm max}}{\left|n_{L,i} - n_{s,i}\right|}\qquad\mbox{,}
\end{equation}
with $n_{L,i}$ being the normalized volume density of large ($L$) particles and $n_{s,i}$ the normalized density of small ($s$) particles at bin $i$. The pre-factor $1/2$ ensures normalization of $P_{\rm segr}$ and $h_{\rm max}$ is the maximum height of the sample. A higher value of $P_{\rm segr}$ indicates a higher degree of segregation. For the mixtures depicted in figure~\ref{fig:density_tomos} we obtain $P_{segr} = 0.06$ for 50\,\%RH and 0.32 at 100\,\%RH. 

To study the impact of charging on segregation in more detail, we conducted segregation experiments in a PTFE cup (diameter 50\,mm) at low ($<\,30\,\%$RH), medium ($30 -\,70\,\%$RH) and high ($>\,70\,\%$RH) humidity using equal-volume mixtures of PA and PTFE grains which differ in their composition. Beads were again shaken for one hour at 100\,Hz and 2\,g and subsequently the reconstructed bead positions allow to compute the vertical volume densities. The experiments were performed three times in each humidity range using the climate chamber described in Sec.~\ref{subsec:humid_control}. 

Different models account for beds consisting of tribo-charged particles by assuming either homogeneous~\cite{Lu2005,Cheng2014} or heterogeneous~\cite{Yoshimatsu2016} charge distribution along the surface of grains. The latter will have an impact on the interaction of the particles that goes beyond simplistic Coulomb-like interactions~\cite{Siu2015,Qin2016}. However, since we only access the total charge of individual grains, we aim at qualitatively linking segregation to the impact of charge by computing the product $Q_{L} Q_{s}$. $Q_{L}$ and $Q_{s}$ are the mean charge value of the large \textit{resp.} small beads in the binary mixture. The charges were measured by extracting ten large and small beads from each bed after shaking using an anti-static tweezer and depositing them individually in a self-made Faraday cup. 
\begin{figure}[htbp]
\includegraphics[width=1\columnwidth]{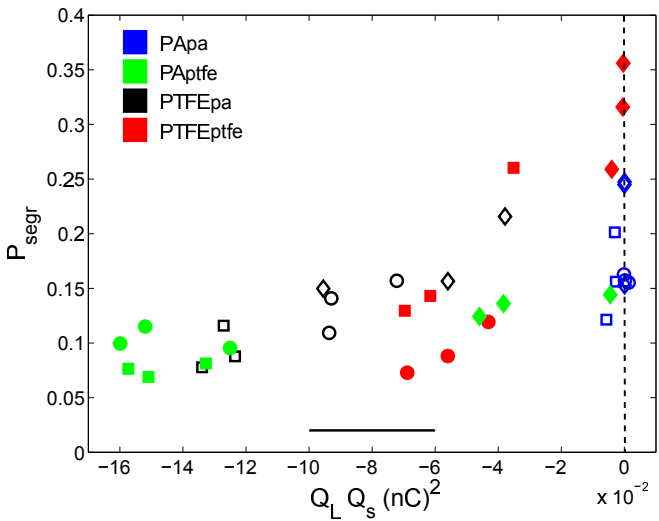}%
	 \caption{Segregation parameter $P_{\rm segr}$ as a function of $Q_{L} Q_{s}$ in binary mixtures. Capital letters denote the large, lower case letters the small component of each mixture. All samples were shaken in a PTFE cup for one hour and at 100\,Hz and 2\,g. Each mixture is coded in a specific color. The symbol shapes indicate the humidity range: squares represent $<30$\,\%RH, circles $30-70$\,\%RH and diamonds $>70$\,\%RH. Low values of $P_{\rm segr}$ are found for larger magnitudes $\mid Q_{L} Q_{s}\mid$. The black bar indicates the overall average uncertainty of the charge measurement.	}%
	 \label{fig:QLQs_vs_segregation}%
\end{figure}

The impact of tribo-charging on granular segregation is shown in figure~\ref{fig:QLQs_vs_segregation} where $Q_{L} Q_{s}$ is plotted against $P_{\rm segr}$. The equal volume mixture consisting of large and small PA particles does not charge when shaken in the PTFE cup. However, the other mixtures do. Even though the data is quite noisy, we can put two statements: First, all products $Q_{L} Q_{s}$ are negative indicating that - to lowest order - large and small particles attract each other. Second, more negative values of $Q_{L} Q_{s}$ always correspond to a lower degree of segregation. Thus, we have shown that electrostatic interactions can not only suppress segregation but also open a venue to create packings by design, cf. Fig.~\ref{fig:granular_crystal}. 

\section{Conclusion}
When samples of polymer particles are shaken vertically in a steel container they do tribo-charge. The sign of the accumulated charge shows some correlation with the water contact angle of the material. The amount can be controlled by the relative humidity of the ambient air. At low humidity levels we observe a plateau of approximately constant charge. Above a material-specific threshold the magnitude of charge decreases to zero or a small level. All these results point to the importance of the water film adsorbed at the surface of the beads for understanding their tribo-electric charge accumulation. Finally, tribo-generated charges can suppress segregation by providing attractive interactions between beads which are strong enough to counteract segregation during shaking. 

\section*{Acknowledgment}\label{sec:acknowledgement}
We acknowledge helpful and inspiring discussions with Scott Waitukaitis, Philip Born, and Victor Lee.
We also want to thank Wolf Keiderling for his support in constructing the automated setup.






\providecommand*{\mcitethebibliography}{\thebibliography}
\csname @ifundefined\endcsname{endmcitethebibliography}
{\let\endmcitethebibliography\endthebibliography}{}

\end{document}